# Light Alkali Metal Functionalized Two-Dimensional $C_5N$ Monolayers for Enhanced Hydrogen Storage

Gom Dorji[1*], Sonam Peden[1], Syed Faraz Hasan[1,2], Kondo-Francois Aguey-Zinsou[3] and Tanveer Hussain[1*]

[1]School of Science and Technology, University of New England, Armidale, New South Wales, 2351, Australia

[2]Directorate of Research Services, University of Canberra, ACT 2617, Australia

[3]MERLin, School of Chemistry, University of Sydney, NSW 2006, Australia

[*]Corresponding authors: gdorji@myune.edu.au, tanveer.hussain@une.edu.au

## Abstract

This work presents a density functional theory (DFT) investigation of a two-dimensional (2D) $C_5N$ monolayer functionalized with Li, Na, and K for hydrogen ($H_2$) storage. Pristine $C_5N$ exhibits weak $H_2$ adsorption, while alkali-metal functionalization significantly enhances its storage capability. The $C_5N$ monolayer can stably accommodate up to six metal dopants, with binding energies stronger than the corresponding cohesive energies, indicating resistance to metal aggregation. Ab initio molecular dynamics simulations further confirm the thermal stability of the functionalized systems at 300 K. Charge transfer from the metal dopants to $C_5N$ enhances polarization and strengthens $H_2$ adsorption. Each dopant can adsorb up to eight $H_2$ molecules, yielding a maximum of 48 $H_2$ molecules per unit cell and gravimetric storage capacities of 9.42, 8.61, and 7.93 wt% for Li-, Na-, and K-functionalized $C_5N$, respectively. The average $H_2$ adsorption energies of −0.16 to −0.17 eV/$H_2$ indicate moderate interactions suitable for reversible storage. Thermodynamic analysis further demonstrates favourable $H_2$ adsorption/desorption under practical operating conditions, while desorption-temperature, recovery-time, and volumetric analyses support the potential reversibility and storage performance of these systems. Overall, alkali-metal-functionalized $C_5N$ emerges as a promising 2D material for efficient and reversible $H_2$ storage.

## 1. Introduction

Global energy consumption is increasing at an average rate of 1-2% per year, with most of the energy still produced from fossil fuels, a major contributor to global greenhouse gas emissions [1]. The transition from fossil fuels to renewable energy (RE) sources is therefore important to achieving net-zero emission targets. However, the intermittency and variability of RE sources, such as solar and wind, pose significant challenges to energy storage, grid stability, and system reliability. Beyond environmental benefits, reducing dependence on fossil fuels can improve public health by mitigating the impacts of climate change and stimulate economic growth through the expansion of green technologies [2-5]. Although the levelized cost of RE has been steadily decreasing, the mismatch between energy supply and demand remains a critical issue [6, 7]. Energy storage systems (ESS) provide a viable solution to this challenge by storing surplus energy during periods of low demand and high generation and delivering it when demand is high or low, or when there is a low renewable output. Among various storage technologies, chemical energy storage, particularly hydrogen ($H_2$), has attracted significant attention due to its light weight, environmental friendliness, natural abundance, high gravimetric energy density (~33 kWh $kg^{-1}$), and its ability to generate energy without $CO_2$ emissions [8-11]. However, efficient and safe $H_2$ storage remains a critical challenge that limits its wider application due to its low density and gaseous nature [12, 13]

The large-scale industrial application of $H_2$ energy is currently constrained by the lack of storage systems that simultaneously meet cost, efficiency, and safety requirements [10, 14]. To address this challenge, extensive research has been conducted on various $H_2$ storage methods, including compressed gas, liquefaction, and solid-state storage systems. Compressed gas storage is the simplest and most economical approach; however, the requirement for high-pressure operation, typically in the range of around 700 bar, necessitates specialized tanks, which raises significant safety concerns related to leakage, material integrity, and explosion risks [15, 16]. Liquid $H_2$ storage, widely utilized in space applications, offers higher density but requires cryogenic conditions (around 20 K, -253 °C), which significantly reduces overall system efficiency due to high liquefaction energy demands, boil-off losses, and challenges associated with thermal insulation and material stability [17].

In contrast to the conventional storage methods, solid-state $H_2$ storage has emerged as a promising alternative due to its potential for higher storage capacity and safer operation. This approach relies on the adsorption of $H_2$ with the host materials, either physisorption or chemisorption mechanisms. Materials such as metal-organic frameworks (MOFs) and covalent organic frameworks (COFs) have been extensively studied due to their high surface area, tunable structures, and porous nature, which provide abundant adsorption sites for $H_2$ molecules [18, 19]. However, $H_2$ adsorption in these materials is governed by weak van der Waals interactions, leading to poor retention at moderate temperatures [20]. Alternatively, metal hydrides (MHs) offer compact and efficient storage through chemisorption, where $H_2$ molecules dissociate and form metal-hydrogen (M-H) bonds. Although MHs exhibit higher gravimetric density compared to MOFs, their practical application is hindered by slow desorption kinetics and poor reversibility [21-23].

The ideal storage must possess the $H_2$ binding energies between physisorption and chemisorption. Recent studies have explored solid-state $H_2$ storage using carbon-based nanostructures such as carbon nanotubes (CNTs) and graphene, particularly when decorated with transition, alkali, or alkaline-earth metals [24]. However, CNTs, despite their potential, transition metal doping often leads to metal clustering, which reduces the number of active adsorption sites [25]. Consequently, two-dimensional (2D) materials, including graphene, MXenes, hexagonal boron nitride (h-BN), borophene, and phosphorene, have attracted significant attention due to their unique electronic, mechanical, and chemical properties [26-28].

Among various options, carbon nitride ($C_xN_y$) has emerged as a promising 2D material owing to its low density, high porosity, large surface area, structural tunability, and chemical stability. It has demonstrated potential in applications such as gas purification, optoelectronics, nano sensors, and $H_2$ storage [29]. It has been reported that the $C_2N$ structure can achieve a $H_2$ storage capacity of 11.62 wt% at a desorption temperature of 229.23 K, with stability confirmed through ab initio molecular dynamics (AIMD) simulations [30]. A more recent work has focused on enhancing $H_2$ storage performance by decorating $C_xN_y$ materials with metal dopants. For instance, lithium-decorated $C_2N$ exhibits $H_2$ storage capacity of 13 wt% with adsorption energies ($E_{ads}$) ranging from -0.12 to -0.23 eV/$H_2$ [31], while Mg-decorated $C_3N_4$ achieves 7.96 wt% with $E_{ads}$ between -0.10 to -0.25 eV/$H_2$ [32]. Additionally, combine decoration of Mg and Li on g-$C_3N_4$ results in a storage capacity of 10.01 wt% and an average $E_{ads}$ of -0.128 eV/$H_2$, with slight elongation of the $H_2$ bond (~0.75 Å) [33]. Similarly, Li-decorated graphene-like $C_{10}N_3$ demonstrates strong metal anchoring (binding energy of -3.37 eV/atom) and a storage capacity of 8.0 wt%, surpassing the U.S. DOE targets, with favourable $E_{ads}$ ranging from -0.208 to -0.227 eV/$H_2$ [34, 35]. Experimental studies, such as Pd-decorated g-$C_3N_4$, report a $H_2$ storage capacity of 2.60 wt% [36].

In this context, our previous work demonstrated $H_2$ storage capacities of 9.47, 5.96, and 6.57 wt% for 4Mg-$C_3N_2$, 4K-$C_3N_2$, and 4Ca-$C_3N_2$ systems, respectively, under fuel cell conditions using DFT calculations, with $E_{ads}$ a range from -0.15 to -0.60 eV/$H_2$ [37]. Similarly, Chen et al. reported that Li-decorated $C_3N_2$ can adsorb up to 12$H_2$ molecules with an average adsorption energy of -0.228 eV/$H_2$ [38].

Motivated by the potential of $C_xN_y$ for $H_2$ storage, in this work, we investigate atomically thin networks based on $C_5N$, which has recently emerged as a new class of organic 2D materials, distinguished by a fused aromatic topology unlike that of traditional inorganic counterparts [39]. This framework supports highly efficient charge transport, with reported carrier mobilities approaching 996 $cm^2\ V^{-1}\ s^{-1}$ for electrons and 501 $cm^2\ V^{-1}\ s^{-1}$ for holes, values that surpass most organic semiconductors. Its direct bandgap (~2.63 eV) and strong absorption in the visible region further broaden its suitability for optoelectronic and photocatalytic applications [39-41]. Mechanically, the material combines high tensile strength (>10 GPa) with moderate lattice thermal conductivity (~9.5 $Wm^{-1}K^{-1}$) and notable resistance to oxidative degradation. Collectively, these features, large accessible surface, efficient charge transport, tunable heat conduction, and structural resilience, differentiate $C_5N$ from established carbon-nitrogen systems. Building on these attributes, we examine alkali metal-functionalized $C_5N$ monolayers (Li, Na, and K) for reversible $H_2$ storage. Decorating the surface with light alkali metals induces charge redistribution and polarization effects that enhance interactions with $H_2$ molecules. The interplay between the intrinsic electronic structure of $C_5N$ and metal-induced modifications enables control over adsorption strength and release conditions. This strategy offers a pathway toward designing stable, high-performance $H_2$ storage materials with advantages in tunability and scalability compared with conventional inorganic systems. It is worth mentioning that the selection of these dopants is motivated by their natural abundance, low atomic weight, and favourable electronic properties. These metals act as electron donors and exhibit strong binding with the substrate, with binding energies exceeding their cohesive energies, thereby preventing metal clustering and ensuring uniform dispersion. Such uniformity is essential for maximizing adsorption sites and enhancing gravimetric storage capacity [42, 43].

## 2. Computational methodology

DFT calculations were performed using the generalized gradient approximation (GGA) developed by the Perdew-Burke-Ernzerhof (PBE) exchange-correlation functional as implemented in the Vienna Ab-initio Simulation Package (VASP) [44-47]. The projector augmented-wave (PAW) method was employed to describe the interaction between valence electrons and ionic cores [48]. A plane-wave cut-off energy was set to 500 eV to ensure computational accuracy. The structure optimization was carried out using the conjugate gradient algorithm until the Hellmann-Feynman forces on each atom were reduced below 0.05 eV/Å with an electronic energy convergence criterion of $10^{-6}$ eV. The van der Waals interactions were considered using the Grimme DFT-D3 correction method [49-52]. The Brillouin zone was sampled using Monkhorst-Pack k-point grids of 3x3x1 for structural optimization [40] and denser k-points of 6x6x1 for electronic property calculations [41, 53]. To prevent interaction of periodic unit cells with adjacent layers, a vacuum of ~20 Å was applied along the z-axis. The Bader charge analysis was conducted to gain insight into charge transfer [54, 55]. The ab initio molecular dynamics (AIMD) simulation of metal-decorated $C_5N$ was performed using a two-step process. Initially, the system was gradually heated from 0 to 300 K over 5 ps under the NVE ensemble with a time step of 1 fs. The system was then equilibrated at 300 K for another 5 ps under the NVT ensemble employing Nose-Hoover thermostat [56].

The binding energy ($E_b$) to determine the structural stability of the metal-doped system is calculated using Eq. (1). The terms $E_{(xM+C5N)}$, $E_{(C5N)}$, $E_m$ and '$x$' represents the total energy of metal-doped $C_5N$, the energy of pristine $C_5N$, the energy of isolated metal dopant, and '$x$' is the concentration of dopants, respectively.

$$E_b = \frac{E_{(xM+C_5N)} - E_{C_5N} - xE_{(M)}}{x} \qquad (1)$$

The charge transfer between dopant and host material is analysed using Eq. (2). The terms $\Delta\rho$ represents charge density difference, $\rho_{(6M+C5N)}$ gives total charge of the metal-doped system, $\rho_{(C5N)}$ represents charge density of pristine system and $\rho_{(M)}$ represents total charge of the metal dopant.

$$\Delta\rho = \rho_{(6M+C_5N)} - \rho_{(C_5N)} - 6\rho_{(M)} \quad (2)$$

The $H_2$ adsorption energy ($E_{ads}$) is calculated using Eq.(3) [37]. The right-hand sides of the first, second, and third terms of Eq. (3) represent energies of the hydrogenated system, total energies of the system with metal dopants, and the energies of isolated $H_2$ molecules, respectively. The *n* represents the number of $H_2$ molecules taken up in the system.

$$E_{ads} = \frac{E_{(C_5N+6M+nH_2)} - E_{(C_5N+6M)} - nE_{H_2}}{n} \quad (3)$$

To ensure system efficiency, stability, and match $H_2$ release rates at fuel cell conditions, the desorption temperature ($T_d$) is calculated using the Van't Hoff equation given by Eq. (4) [57]. The terms $K_B$ is the Boltzmann constant, R is the universal gas constant (8.314 J $mol^{-1}K^{-1}$), and ΔS is the entropy change of 75.44 J $mol^{-1}K^{-1}$ adopted from the reported value for $H_2$ adsorption on surfaces [52, 58, 59].  P signifies the equilibrium pressure specified as 1 atm [10]

$$T_d = \frac{|E_{ads}|}{K_B\left(\frac{\Delta S}{R} - \ln P\right)} \quad (4)$$

The recovery time (τ) is calculated using Eq$\tau = \frac{1}{\omega} e^{\frac{|E_{ads}|}{K_B T}}$ (5. (5) The ω and T denote the attempt frequency 1x$10^{-12}$ s and thermodynamic temperature at 298.15 K, respectively [52].

$$\tau = \frac{1}{\omega} e^{\frac{|E_{ads}|}{K_B T}} \quad (5)$$

Finally, the effectiveness of pristine $C_5N$ for $H_2$ storage capacity is calculated using Eq. (6), where $nH_2$ is the total molecular mass of $H_2$ and $MC_5N$ is the total molecular mass of the metal-decorated monolayer.

$$wt\% = \frac{nH_2}{nH_2 + 6MC_5N} x100 \quad (6)$$

To represent the real-world scenario, the average number of $H_2$ molecules adsorption and desorption on pristine $C_5N$ was analysed by thermodynamic characteristics. The effective number of $H_2$ molecules ($N_{eff}$) stored in the host materials as a function of variable temperature (T) and pressure (P) is given by Eq. (7).

$$N_{eff}(T,P) = \left[\frac{Z-1}{Z}\right] * N_o \quad (7)$$

Where Z defines the grand canonical partition function as shown in Eq.(8)$Z(E_{ads}, T, P) = 1 + \sum_{i=1}^{n} e^{-\left(\frac{E^i_{ads} - \mu_{H_2}}{K_B T}\right)}$ (8), $N_o$ represents the number of adsorbed $H_2$ molecules at 0.0 K calculated via VASP simulation.

$$Z(E_{ads},T,P) = 1 + \sum_{i=1}^{n} e^{-\left(\frac{E_{ads}^{i}-\mu_{H_2}}{K_B T}\right)} \quad (8)$$

Where $E_{ads}^{i}$ is the average adsorption energy of i[th] $H_2$ molecule, and 'i' index starts from 12, 24, 36, 42, and 48. The 'n' defines the upper limits of $H_2$ molecules stored in decorated $C_5N$ with $E_{ads}$ greater than -0.15 eV/$H_2$. Each metal atom could successfully adsorb 8 $H_2$ molecules. This $\mu_{H_2}$ is the gas-phase chemical potential calculated using Eq.(9).

$$\mu_{H_2}(P,T) = \Delta H(T,P) - T\Delta S(T,P) + K_B T \ln\frac{P}{P_o} \quad (9)$$

Where $\Delta H - T\Delta S$ is Gibbs free energy calculated using the Shomate equation from [60]. ΔH is the standard enthalpy change of $H_2$ in kJ $mol^{-1}$, calculated using Eq. S1 (supplementary equation 1), ΔS is the standard entropy change calculated using Eq. S2 (supplementary equation 2) in $Jmol^{-1}K^{-1}$, both referenced to 298.15 K.

# 3. Results and discussion

## 3.1 Structural and electronic properties of $C_5N$

The pristine $C_5N$ monolayer consists of 60 carbon (C), 12 nitrogen (N), and 18 hydrogen (H) atoms in a hexagonal unit cell with α=β=90° and γ=120°. The 18H atoms are used to passivate the dangling bonds when the unit cell is formed. It should be noted that these H atoms are not accounted for in any $H_2$ storage calculations.

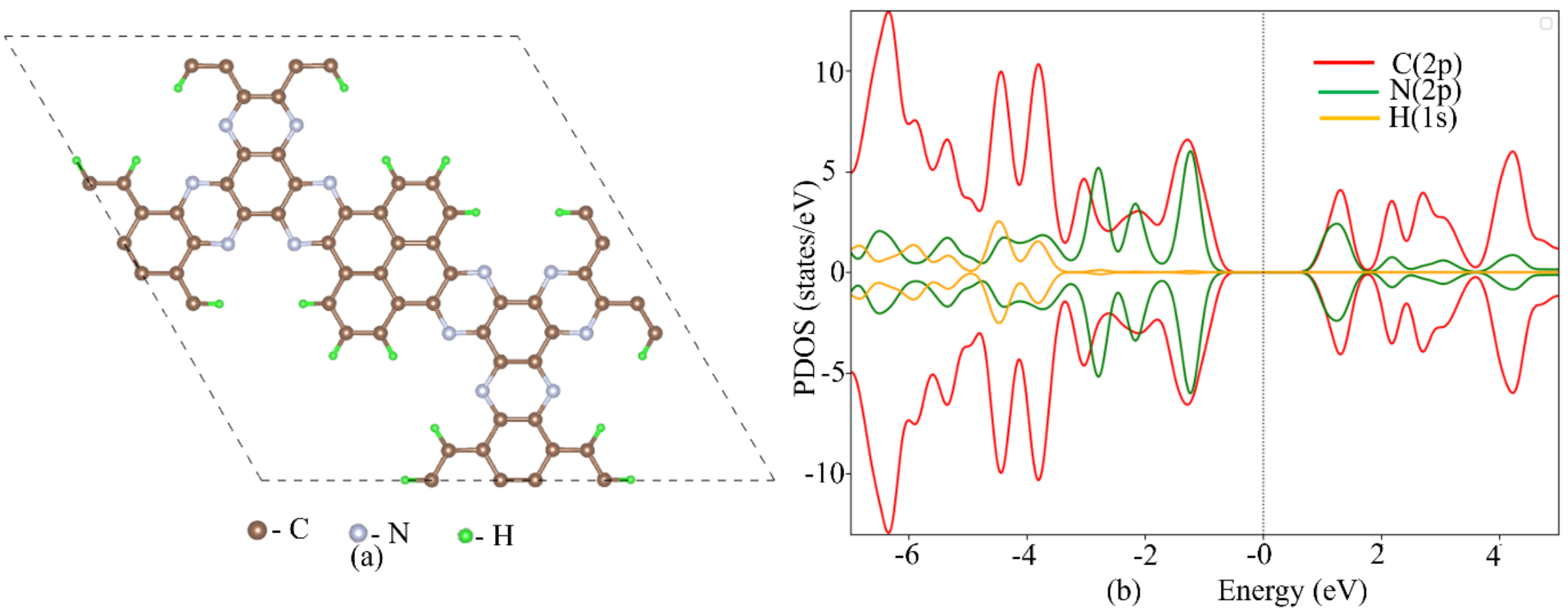

Figure 1. (a) Optimized structure of $C_5N$ (brown – carbon, grey – nitrogen, green-hydrogen). (b) Electronic properties showing orbital hybridization and band gaps.

Figure 1 (a) shows the optimized structure of $C_5N$. There are three types of bonds. The bond length between C-C ranges from 1.39 -1.47 Å, C-N has a bond length of 1.34 Å, and C-H has a bond length of 1.09Å, which agrees with the reported values [40, 41]. Figure 1(b) shows the spin-polarised projected density of states (PDOS) of pristine $C_5N$. The region below the Fermi level corresponds to the occupied valence states, where the orbital interactions contributing to bonding can be analysed. In this region, several overlapping PDOS peaks are observed at approximately −0.6 eV, −4.0 eV, and −6.2 eV, where both C(2p) and N(2p) orbitals exhibit significant contributions at the same energy positions. These coincident orbital distributions indicate strong interaction and hybridization between C(2p) and N(2p) states. Although other overlapping regions are also present in the valence band, the above-mentioned energy regions show the most pronounced contributions and are therefore identified as the dominant hybridization regions. Furthermore, the peaks around −4.0 eV and −6.2 eV also contain contributions from passivating H atoms, indicating the involvement of H-derived states in modifying the electronic structure of the C–N framework. The energy gap of 1.70 eV is consistent with a semiconducting nature of $C_5N$ [41]. The conduction band is primarily composed of C(2p) orbitals, facilitating charge accumulation in these states. The symmetric spin-polarized PDOS shows the system is non-magnetic in ground states.

### 3.2 Metal functionalization of $C_5N$

Like most of the 2D materials, pristine $C_5N$ has weak interactions with $H_2$ [41]. To enhance $H_2$ adsorption, selected light metals were decorated on $C_5N$. The metal dopants are introduced at available binding sites, such as on porous, C-top, N-top, inside the C-hexagonal ring, on C-C bonds, and C-N bonds. The most preferential binding site was identified through systematic calculations, and the highest magnitude of $E_b$ was calculated using Eq. (2). The negative sign indicates the process is exothermic.

A homogeneous dispersion of the dopants over the $C_5N$ monolayer is essential for preserving structural integrity and ensuring reversible functional performance. To evaluate the stability of the metal-doped $C_5N$, the magnitude of $E_b$ is compared with cohesive energy ($E_c$) as shown in Figure 2. It is observed that an increase in doping concentration makes binding energy less negative. The doping concentration is also determined by the adsorption sites. From Figure 2, it is observed that $E_b$ are stronger than the corresponding $E_c$, indicating that binding of the dopants on the $C_5N$ is energetically more favourable than metal clustering. This confirms the structural stability of the metal-doped $C_5N$ and suggests a low tendency for dopant aggregation. Among the investigated systems, Li-doped C5N (Li-C5N) exhibits the strongest and most consistent $E_b$ across all doping concentrations, followed by Na, whereas K shows the weakest and most variable binding behaviour. This trend originates from differences in dopant-C5N interactions. The smaller ionic radius of Li promotes stronger orbital overlap with the $C_5N$ and facilitates more effective charge transfer, leading to enhanced stabilization. In contrast, the larger atomic size of K weakens orbital coupling and reduces adsorption stability. The comparatively poorer performance of K-doped $C_5N$ at certain concentrations is likely associated with local lattice strain and weakened metal-substrate interactions. The optimal number of metal dopants accommodated on $C_5N$ was determined through systematic calculations, with six dopants identified as the most stable configuration, as illustrated in Figure 3. The stepwise metal decoration at different concentrations is shown in Figures S1 to S3 (Supporting Information).

The bond length of dopant to dopant on $C_5N$ system is also compared with bulk metal-to-metal bond length, as shown in Table 1. The bond lengths of the metal dopants on $C_5N$ were notably longer than their corresponding bulk metal-metal bond lengths, which further confirms that metal-metal aggression is unlikely. Table 1 provides the numerical data, whereas Figure S4 presents the corresponding visual (graphical) representation showing distance between Li2 and Li3 is 6.47 Å. The calculated distances are also presented in Figure S4 (Supporting Information).

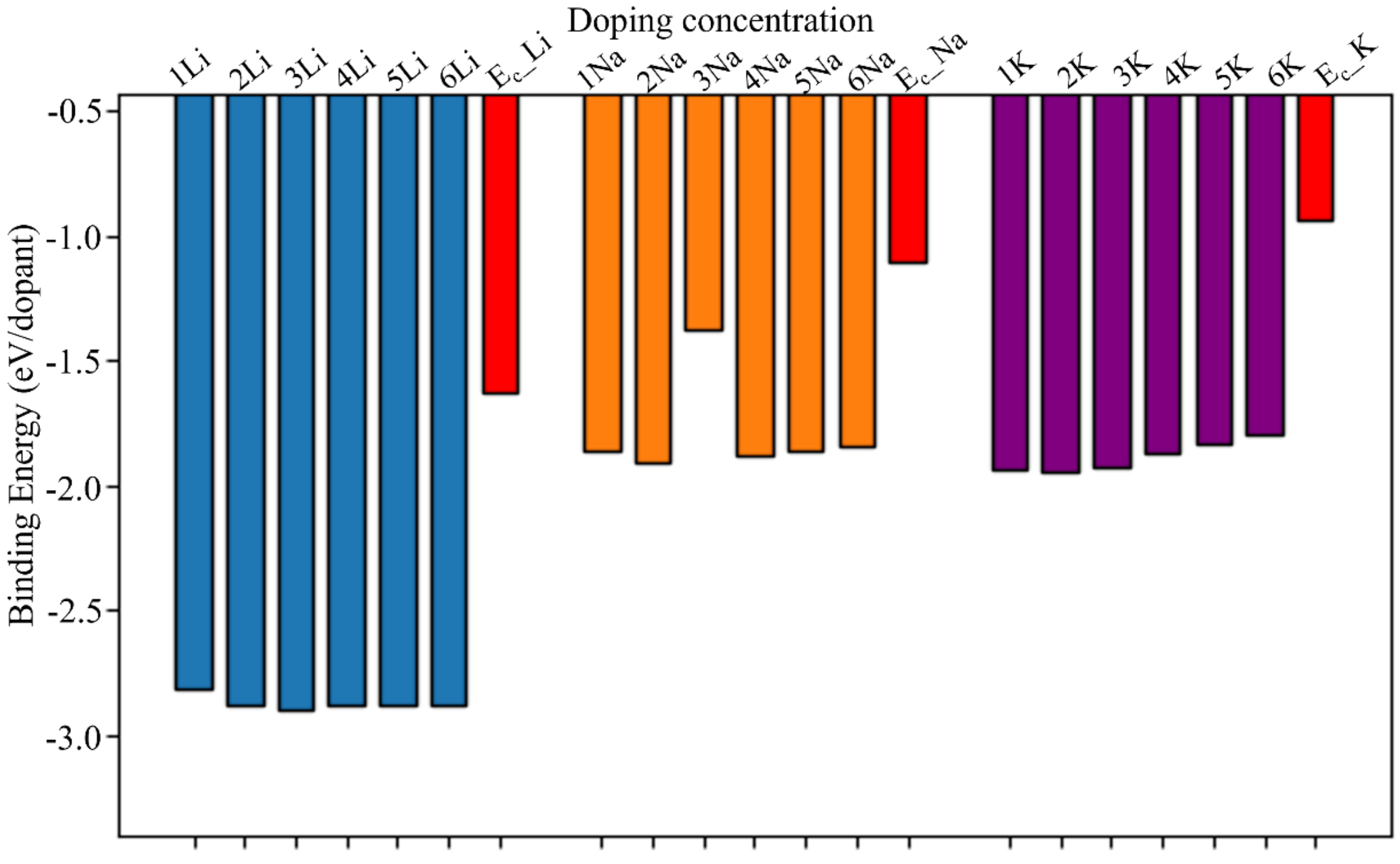

Figure 2. The $E_b$ values of Li, Na, and K on $C_5N$ at different doping concentrations. The magnitude of corresponding $E_c$ values from literature [61] are also plotted and shown in red as $E_c$_Li, $E_c$_Na and $E_c$_K for lithium, sodium and potassium respectively.

Table 1. Comparison of the dopant-dopant distances with the existing literature to confirm the least possibility of metal aggregation.

| Dopants | Inter-dopant distances (Å) | |
|---|---|---|
| | Calculated in this study | Bulk values |
| Li | 6.47 | 2.67 [62] |
| Na | 7.00 | 3.08 [62] |
| K | 7.6 | 3.91 [62] |

Moreover, the thermal stabilities of the metal functionalized $C_5N$ at the maximum doping concentrations (6M-$C_5N$; M= Li, Na, K) were confirmed through AIMD simulation at an elevated temperature. The AIMD simulations were performed using a two-step process: a canonical (NVE) ensemble, where the temperature was varied from 0 to 300 K and NVT ensemble using the Nose-Hoover thermostat at 300 K for 5 picoseconds with a time step of 1.0 femtosecond. It was observed that there are negligible variations in system energy, which indicates that the 6M-$C_5N$ are thermodynamically stable at higher temperatures. The AIMD results are shown in Figure 3.

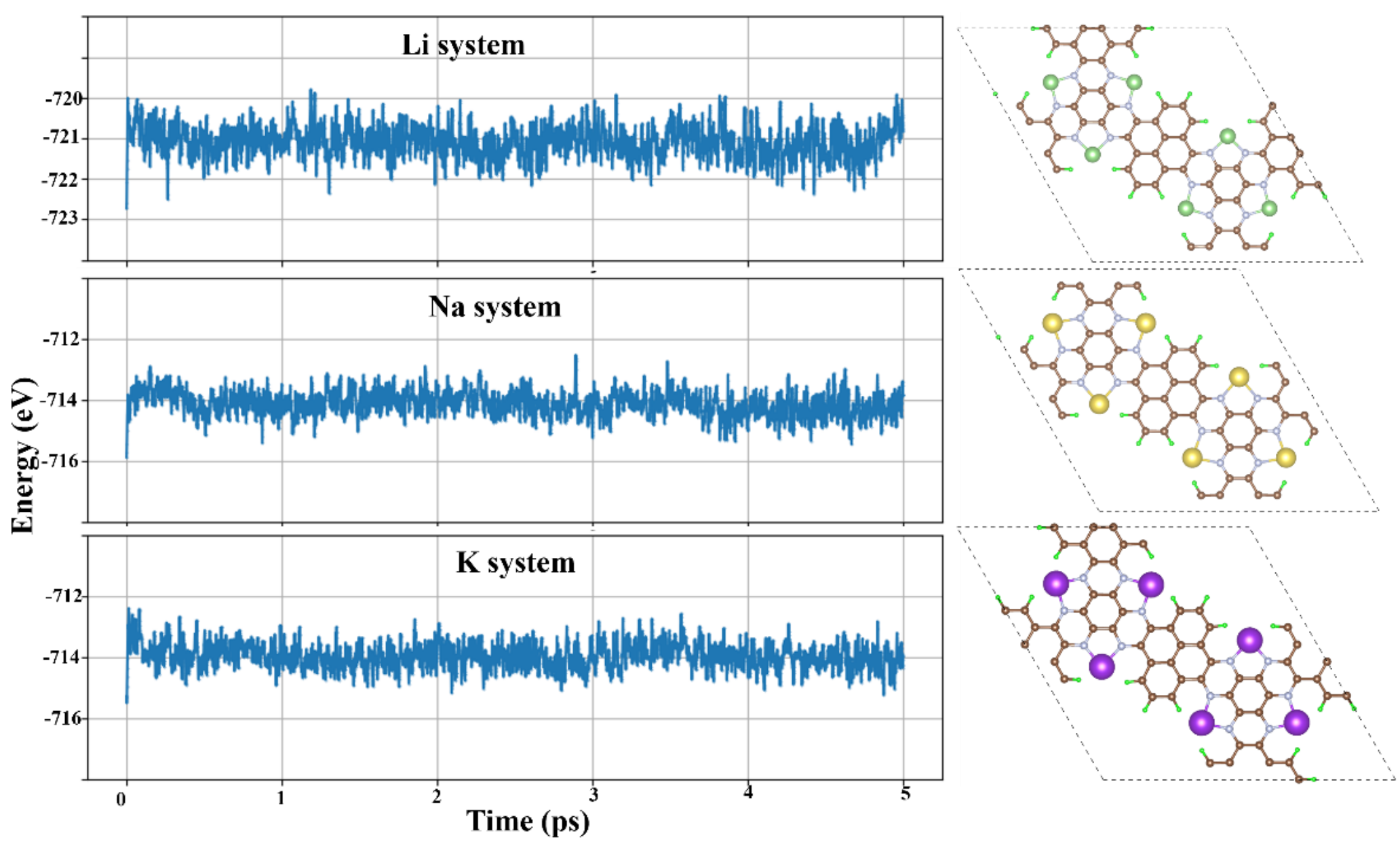

Figure 3. Variation of total energies at an elevated temperature and the optimized structures of 6M-$C_5N$. (light green- Li, yellow – Na, and purple – K)

Next, the electronic properties of the 6M-$C_5N$ system were studied through the projected density of states (PDOS), as shown in Figure 4. All systems exhibit finite electronic states in the vicinity of the Fermi level, confirming their metallic nature. Compared to pristine $C_5N$, metal doping significantly enhances the density of states at the Fermi level, leading to a transition from semiconductor to metallic behaviour. Approximately between -1.80 and -0.80 eV, there exists a clear depletion indicating the presence of a forbidden gap arising from the separation of bonding and antibonding states. In the valence band region, sharp peaks are mainly contributed by the p orbitals of C and N atoms, with minor contributions from s orbitals of H and M atoms, suggesting strong hybridization between the dopant atoms and the pristine $C_5N$, which contributes to structural stability. In the conduction band region, the overlap of C-2p, N-2p, and metals' 's' orbitals suggests strong electronic hybridization, which can facilitate charge transfer and promote $H_2$ adsorption.

The qualitative analysis of charge transfer and electronic redistribution at the atomic level between metal dopants and $C_5N$ is performed through Bader charge calculation. The Bader charge analysis conducted on 6Li-$C_5N$, 6Na-$C_5N$, and 6K-$C_5N$ systems shows that a total of 0.99, 0.99, and 0.85 electrons/dopant, respectively, were transferred to $C_5N$. The metal dopants act as charge donors and $C_5N$ as recipient. The quantitative analysis is conducted through charge density difference using Eq. (2) and is depicted in Figure 5. The overlap of charge accumulation (yellow regions) between C and N atoms indicates the presence of covalent bonding within the $C_5N$ framework. In the 6M-$C_5N$ system, the charge density difference plot shows yellow regions around C and N atoms, indicating charge accumulation, while cyan regions around Li, Na, and K atoms represent charge depletion. This suggests that dopants donate electronic charges to the $C_5N$. As Li, Na, and K readily donate electrons and form positively charged cations, while the C and N atoms of $C_5N$ act as charge acceptors. The resulting charge redistribution creates an electrostatic field that can facilitate $H_2$ polarisation, thereby enhancing $H_2$ adsorption.

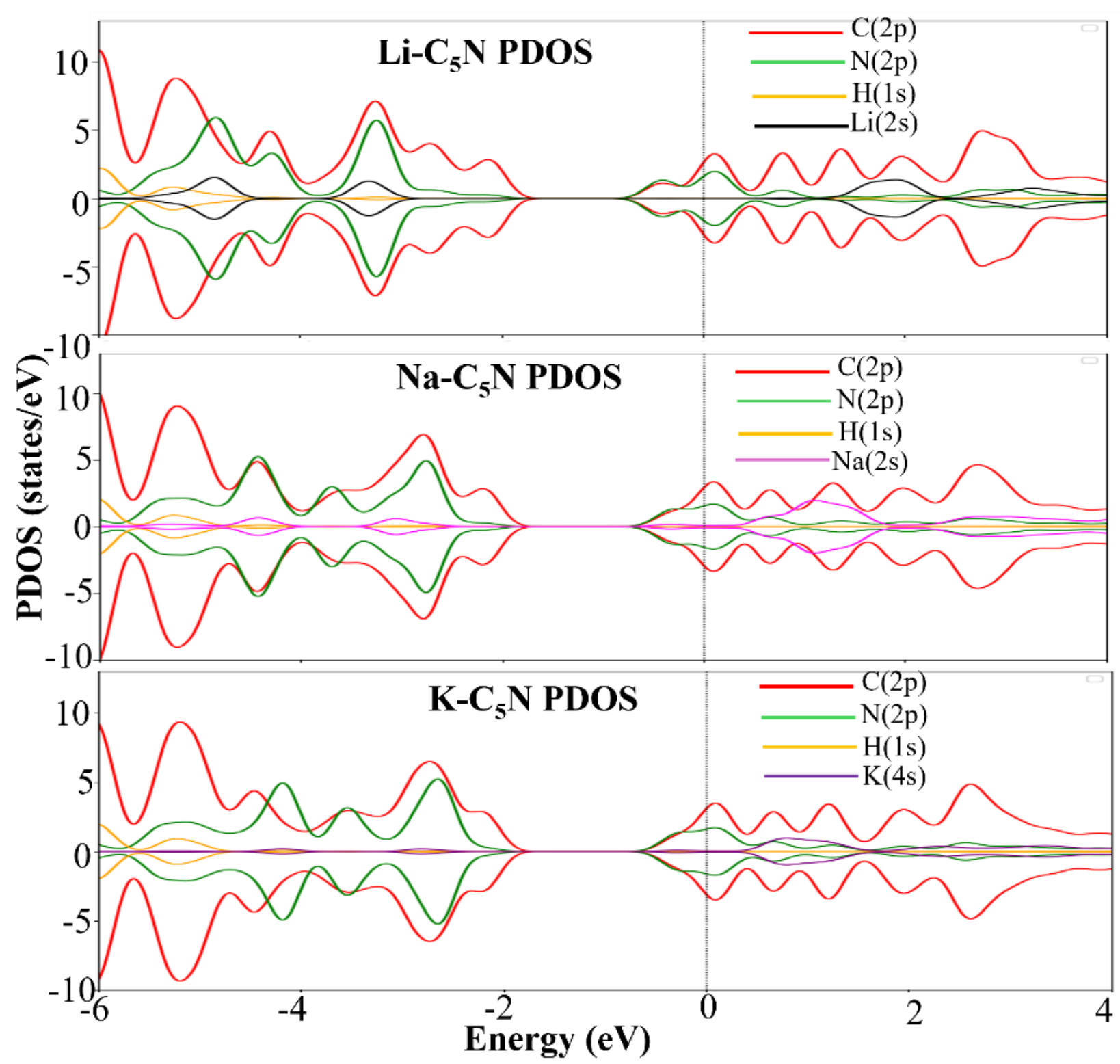

Figure 4. Projected density of state plot for 6Li-$C_5N$, 6Na-$C_5N$, and 6K-$C_5N$. The Fermi level is adjusted to 0.

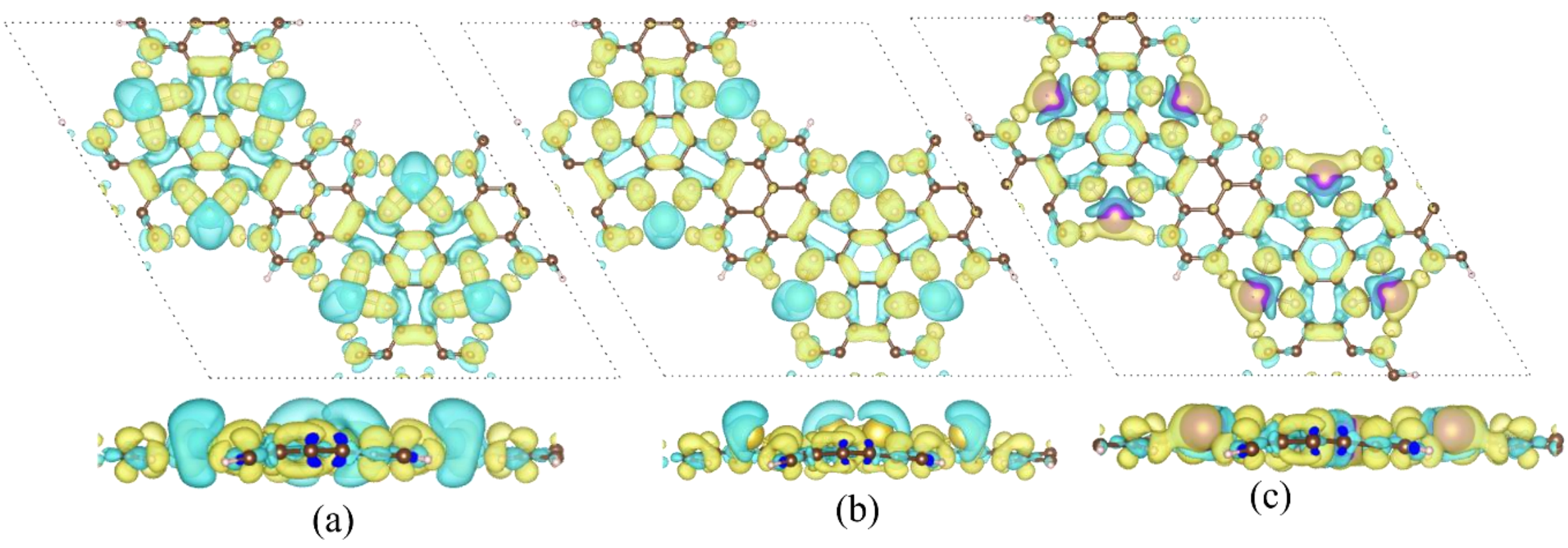

Figure 5. Charge density differences of (a) 6Li-$C_5N$, (b) 6Na-$C_5N$, and 6K-$C_5N$. Cyan colour represents charge depletion, yellow colour represents charge accumulation with iso surface 0.004e/$Å^3$.

### 3.3 Hydrogenation of 6M-$C_5N$

After exploring structural, electronic, charge transfer, and thermal stabilities, the 6M-$C_5N$ monolayers are studied for hydrogenation. To investigate the hydrogen adsorption mechanism, a comprehensive analysis was carried out by examining the adsorption energy, storage capacity, desorption temperature, and adsorption geometry. The $H_2$ molecules were systematically introduced on 6M-$C_5N$, following methodologies widely adopted in the literature for metal-doped 2D materials [35, 38]. Hydrogenation was initiated using $H_2$ molecules positioned above and below each metal dopant site, corresponding to the adsorption of 12 $H_2$ molecules at a time, and the $E_{ads}$ values were calculated using Eq.(3). With an increase in $H_2$ molecule loading, $E_{ads}$ gradually decreases. The distance between 6M-$C_5N$ and $H_2$ molecules was maintained at 1.90 Å, and the H-H was set to 0.74 Å before optimization. During the structural optimization, both H-H bond length and distance of $H_2$ from the 6M-$C_5N$ were elongated and extended to 0.78 Å and 3.5 Å (for maximum hydrogenation), respectively. These short distances of 3.5 Å between $H_2$ molecules and host materials indicate $H_2$ molecule retention via physisorption [37]. The detailed information about the $E_{ads}$, H-H bond length and $H_2$ distance from 6M-$C_5N$ is given in

Table 2. Adsorption energies, $H_2$ distance from host material, H-H bond length after structure optimization, desorption temperature, and recovery time.

| System | $E_{ads}$ (eV) | H-H bond length (Å) | $H_2$ distance (Å) | $T_d$ (1bar) (K) | $T_d$ (10bar) (K) | Recovery time τ (s) |
|---|---|---|---|---|---|---|
| $12H_2$-$6LiC_5N$ | -0.17 | 0.75-0.78 | <3.5 | 217.42 | 347.78 | $7.475x10^{-10}$ |
| $24H_2$-$6LiC_5N$ | -0.17 | 0.75-0.78 | <3.5 | 217.42 | 347.78 | $7.475x10^{-10}$ |
| $36H_2$-$6LiC_5N$ | -0.16 | 0.75-0.78 | <3.5 | 204.63 | 327.32 | $5.065x10^{-10}$ |
| $42H_2$-$6LiC_5N$ | -0.16 | 0.75-0.78 | <3.5 | 204.63 | 327.32 | $5.065x10^{-10}$ |
| $48H_2$-$6LiC_5N$ | -0.16 | 0.75-0.78 | <3.5 | 204.63 | 327.32 | $5.065x10^{-10}$ |
| $12H_2$-$6NaC_5N$ | -0.17 | 0.75-0.78 | <3.5 | 217.42 | 347.78 | $7.475x10^{-10}$ |
| $24H_2$-$6NaC_5N$ | -0.17 | 0.75-0.78 | <3.5 | 217.42 | 347.78 | $7.475x10^{-10}$ |
| $36H_2$-$6NaC_5N$ | -0.17 | 0.75-0.78 | <3.5 | 217.42 | 347.78 | $7.475x10^{-10}$ |
| $42H_2$-$6NaC_5N$ | -0.16 | 0.75-0.78 | <3.5 | 204.63 | 327.32 | $5.065x10^{-10}$ |
| $48H_2$-$6NaC_5N$ | -0.17 | 0.75-0.78 | <3.5 | 217.42 | 347.78 | $7.475x10^{-10}$ |
| $12H_2$-$6KC_5N$ | -0.16 | 0.75-0.78 | <3.5 | 204.63 | 327.32 | $5.065x10^{-10}$ |
| $24H_2$-$6KC_5N$ | -0.16 | 0.75-0.78 | <3.5 | 204.63 | 327.32 | $5.065x10^{-10}$ |
| $36H_2$-$6KC_5N$ | -0.17 | 0.75-0.78 | <3.5 | 217.42 | 347.78 | $7.475x10^{-10}$ |
| $42H_2$-$6KC_5N$ | -0.17 | 0.75-0.78 | <3.5 | 217.42 | 347.78 | $7.475x10^{-10}$ |
| $48H_2$-$6KC_5N$ | -0.17 | 0.75-0.78 | <3.5 | 217.42 | 347.78 | $7.475x10^{-10}$ |

It is worth mentioning that the average $E_{ads}$ (-0.16 to -0.17 eV/$H_2$) is slightly weaker than the ideal values (-0.20 eV/$H_2$) for ambient-condition $H_2$ storage. However, the calculated $E_{ads}$ values fall within the moderate range (-0.10 to -0.30 eV/$H_2$) [59, 63, 64], which have been reported for alkali-metal-decorated 2D materials, supporting the validity of the present results [52, 59, 63-66]. Such binding strengths ensure efficient adsorption while still allowing facile desorption, a critical requirement for practical applications. Each system can accommodate up to 48 molecules of $H_2$. The gravimetric density (wt%) of $H_2$ is calculated using Eq.(6). The result shows gravimetric densities of 9.42, 8.61, and 7.93 wt% for 6Li-$C_5N$, 6Na-$C_5N$, and 6K-$C_5N$, respectively, surpassing the US DOE target [67].

The Grand Canonical thermodynamic analysis further confirms that 6M-$C_5N$ can achieve stable $H_2$ uptake under practical temperature and pressure conditions of the fuel cell for real-world application. Using Eq. (7), (8), and (9), thermodynamic analysis was performed to calculate the effective number of $H_2$ molecules uptake under changing temperature and pressure. In all three systems, the $H_2$ uptakes range from 45 to 48 molecules at the temperature range of 280 to 400 K and pressure range of 0 to 50 bar. To consider the fuel cell conditions at the pressure of 5 to 12 bar, the Figure 7 plot shows $H_2$ intake gradually decreasing after the temperature crosses 340 K [41]. At 298.15 K and 10 bar, the storage capacity remains nearly the same.

For practical reversible $H_2$ storage applications, it is equally important to investigate the desorption behaviour of $H_2$ at different temperatures. The $T_d$ value was calculated using Eq.

$$T_d = \frac{|E_{ads}|}{K_B\left(\frac{\Delta S}{R} - ln\, P\right)}$$

(4) under two pressure conditions, namely 1 bar and 10 bar. At 1 bar, the calculated $T_d$ values range from 204 to 217 K, whereas at 10 bar they increase to 327-347 K, indicating that $H_2$ can be effectively released from the $C_5N$ system at elevated temperatures. This behaviour concludes that to store the $H_2$, temperature and pressure should be inversely proportional. The desorption kinetics are studied using Eq.(5). The calculated recovery time indicates that the adsorbed $H_2$ molecules can be released rapidly under practical operating conditions, demonstrating the excellent reversibility of the proposed hydrogen storage system. It is observed that to release the 12 $H_2$ molecules, it takes 0.51 to 0.75 ns.

. The top and side views of the optimized hydrogenated 6M-$C_5N$ structures are shown in Figure 6. The stepwise loading of $H_2$ on 6Li-$C_5N$ is shown in Figure S5 (Supporting Information).

Table 2. Adsorption energies, $H_2$ distance from host material, H-H bond length after structure optimization, desorption temperature, and recovery time.

| System | $E_{ads}$ (eV) | H-H bond length (Å) | $H_2$ distance (Å) | $T_d$ (1bar) (K) | $T_d$ (10bar) (K) | Recovery time τ (s) |
|---|---|---|---|---|---|---|
| $12H_2$-$6LiC_5N$ | -0.17 | 0.75-0.78 | <3.5 | 217.42 | 347.78 | $7.475x10^{-10}$ |
| $24H_2$-$6LiC_5N$ | -0.17 | 0.75-0.78 | <3.5 | 217.42 | 347.78 | $7.475x10^{-10}$ |
| $36H_2$-$6LiC_5N$ | -0.16 | 0.75-0.78 | <3.5 | 204.63 | 327.32 | $5.065x10^{-10}$ |
| $42H_2$-$6LiC_5N$ | -0.16 | 0.75-0.78 | <3.5 | 204.63 | 327.32 | $5.065x10^{-10}$ |
| $48H_2$-$6LiC_5N$ | -0.16 | 0.75-0.78 | <3.5 | 204.63 | 327.32 | $5.065x10^{-10}$ |
| $12H_2$-$6NaC_5N$ | -0.17 | 0.75-0.78 | <3.5 | 217.42 | 347.78 | $7.475x10^{-10}$ |
| $24H_2$-$6NaC_5N$ | -0.17 | 0.75-0.78 | <3.5 | 217.42 | 347.78 | $7.475x10^{-10}$ |
| $36H_2$-$6NaC_5N$ | -0.17 | 0.75-0.78 | <3.5 | 217.42 | 347.78 | $7.475x10^{-10}$ |
| $42H_2$-$6NaC_5N$ | -0.16 | 0.75-0.78 | <3.5 | 204.63 | 327.32 | $5.065x10^{-10}$ |
| $48H_2$-$6NaC_5N$ | -0.17 | 0.75-0.78 | <3.5 | 217.42 | 347.78 | $7.475x10^{-10}$ |
| $12H_2$-$6KC_5N$ | -0.16 | 0.75-0.78 | <3.5 | 204.63 | 327.32 | $5.065x10^{-10}$ |
| $24H_2$-$6KC_5N$ | -0.16 | 0.75-0.78 | <3.5 | 204.63 | 327.32 | $5.065x10^{-10}$ |
| $36H_2$-$6KC_5N$ | -0.17 | 0.75-0.78 | <3.5 | 217.42 | 347.78 | $7.475x10^{-10}$ |
| $42H_2$-$6KC_5N$ | -0.17 | 0.75-0.78 | <3.5 | 217.42 | 347.78 | $7.475x10^{-10}$ |
| $48H_2$-$6KC_5N$ | -0.17 | 0.75-0.78 | <3.5 | 217.42 | 347.78 | $7.475x10^{-10}$ |

It is worth mentioning that the average $E_{ads}$ (-0.16 to -0.17 eV/$H_2$) is slightly weaker than the ideal values (-0.20 eV/$H_2$) for ambient-condition $H_2$ storage. However, the calculated $E_{ads}$ values fall within the moderate range (-0.10 to -0.30 eV/$H_2$) [59, 63, 64], which have been reported for alkali-metal-decorated 2D materials, supporting the validity of the present results [52, 59, 63-66]. Such binding strengths ensure efficient adsorption while still allowing facile desorption, a critical requirement for practical applications. Each system can accommodate up to 48 molecules of $H_2$. The gravimetric density (wt%) of $H_2$ is calculated using Eq.(6). The result shows gravimetric densities of 9.42, 8.61, and 7.93 wt% for 6Li-$C_5N$, 6Na-$C_5N$, and 6K-$C_5N$, respectively, surpassing the US DOE target [67].

The Grand Canonical thermodynamic analysis further confirms that 6M-$C_5N$ can achieve stable $H_2$ uptake under practical temperature and pressure conditions of the fuel cell for real-world application. Using Eq. (7), (8), and (9), thermodynamic analysis was performed to calculate the effective number of $H_2$ molecules uptake under changing temperature and pressure. In all three systems, the $H_2$ uptakes range from 45 to 48 molecules at the temperature range of 280 to 400 K and pressure range of 0 to 50 bar. To consider the fuel cell conditions at the pressure of 5 to 12 bar, the Figure 7 plot shows $H_2$ intake gradually decreasing after the temperature crosses 340 K [41]. At 298.15 K and 10 bar, the storage capacity remains nearly the same.

For practical reversible $H_2$ storage applications, it is equally important to investigate the desorption behaviour of $H_2$ at different temperatures. The $T_d$ value was calculated using Eq. $T_d = \frac{|E_{ads}|}{K_B\left(\frac{\Delta S}{R} - \ln P\right)}$ (4) under two pressure conditions, namely 1 bar and 10 bar. At 1 bar, the calculated $T_d$ values range from 204 to 217 K, whereas at 10 bar they increase to 327-347 K, indicating that $H_2$ can be effectively released from the $C_5N$ system at elevated temperatures. This behaviour concludes that to store the $H_2$, temperature and pressure should be inversely proportional. The desorption kinetics are studied using Eq.(5). The calculated recovery time indicates that the adsorbed $H_2$ molecules can be released rapidly under practical operating conditions, demonstrating the excellent reversibility of the proposed hydrogen storage system. It is observed that to release the 12 $H_2$ molecules, it takes 0.51 to 0.75 ns.

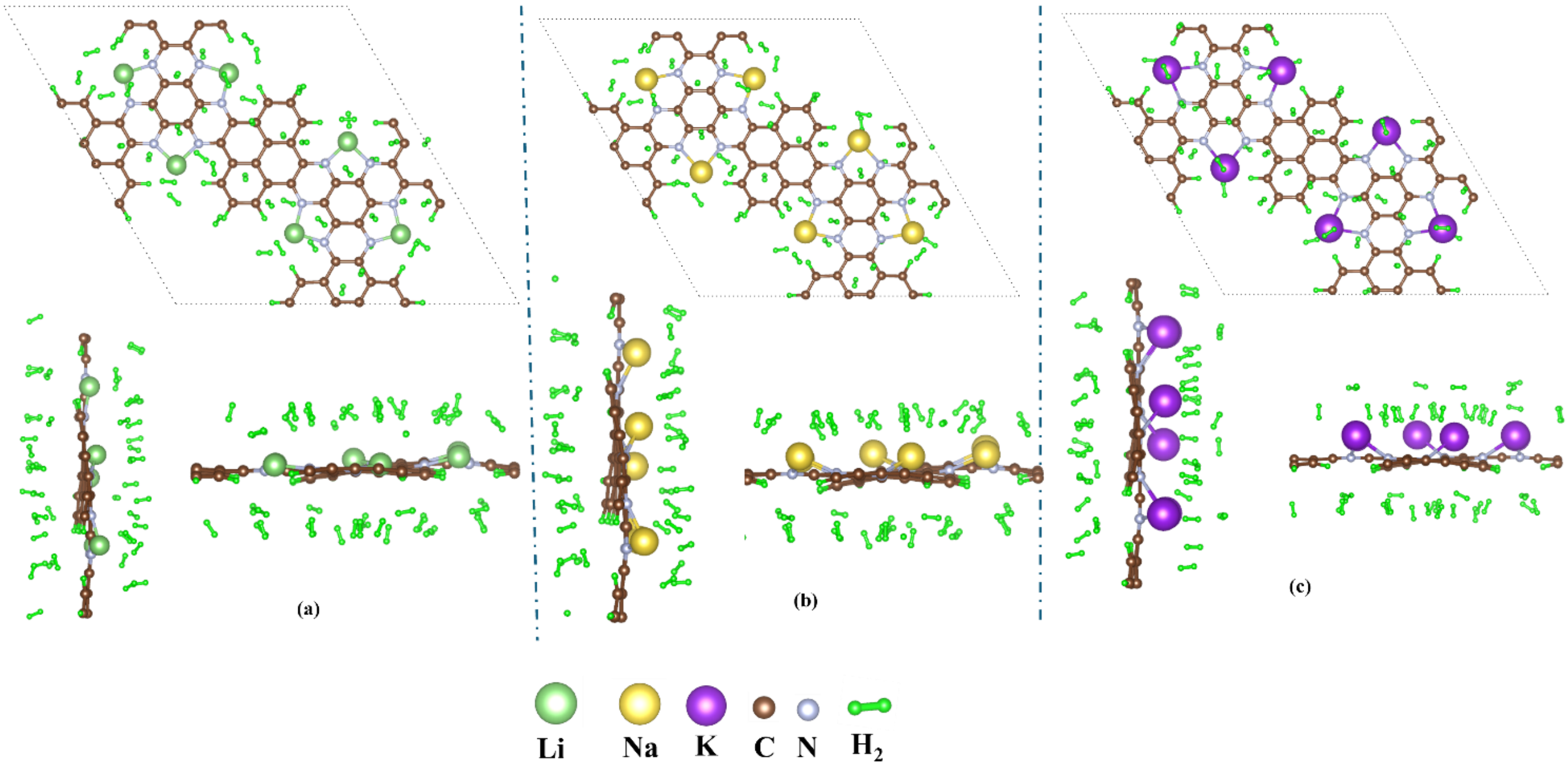

Figure 6. Top and side views of the optimized structures of $H_2$ adsorbed (a) 6Li-$C_5N$, (b) 6Na-$C_5N$, and 6K-$C_5N$ systems. Light green, yellow, purple, brown, grey, and green colours represent atoms of lithium, sodium, potassium, carbon, nitrogen and $H_2$ molecules, respectively.

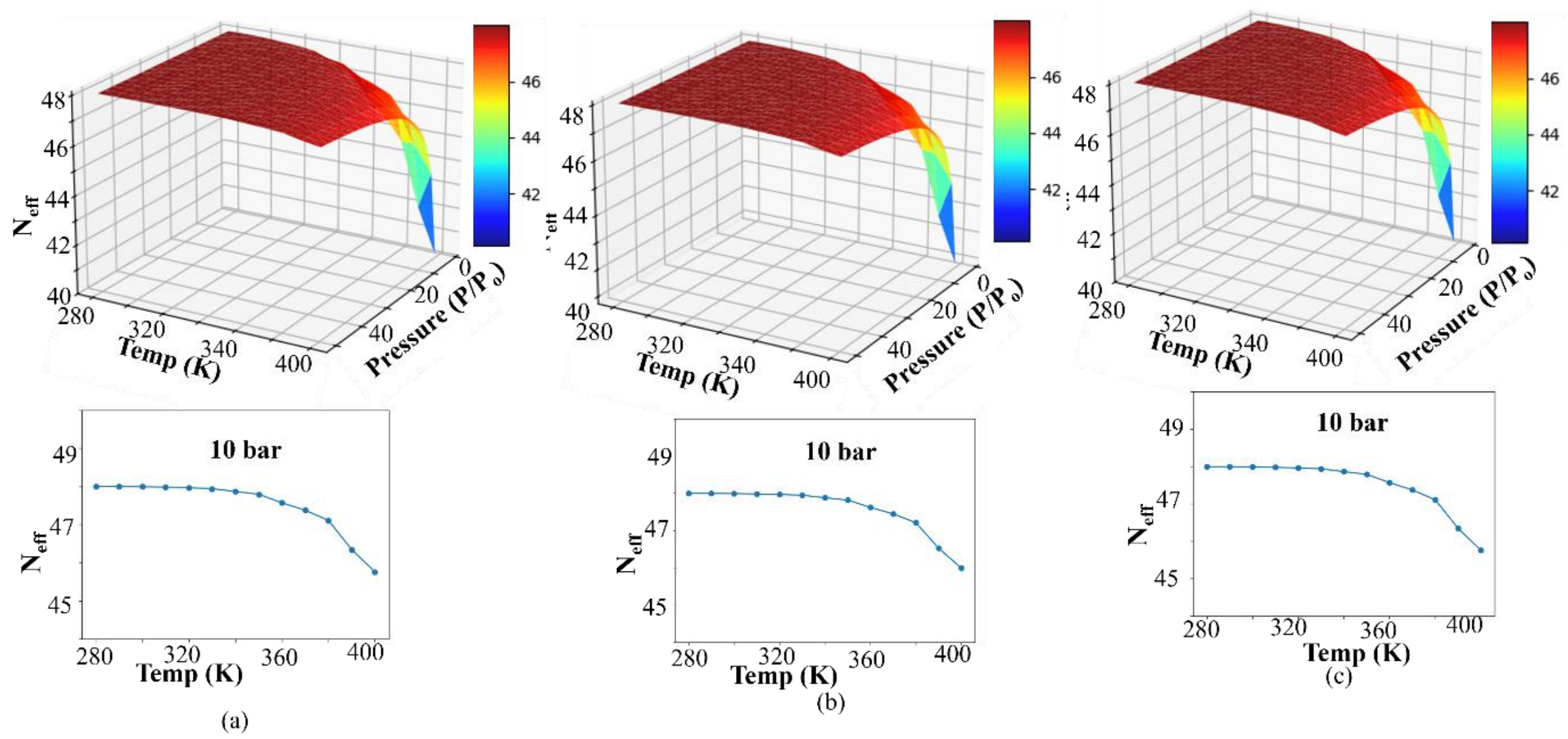

Figure 7. Average number of $H_2$ update on (a) $6Li\text{-}C_5N$, (b) $6Na\text{-}C_5N$, and $6K\text{-}C_5N$ at varying temperature and pressure. The bottom 2D plot shows $H_2$ uptake at a fixed pressure of 10 bar.

The $H_2$ storage capacities of the $6M\text{-}C_5N$ systems were compared with previously reported 2D materials in terms of $E_{ads}$ and the maximum number of adsorbed $H_2$ molecules. Existing reports on transition-metal-decorated $C_5N$ indicate adsorption energies within the DOE-recommended range, with storage capacities only slightly exceeding the target value of 5.5 wt%. In contrast, the 6Li-, 6Na-, and $6K\text{-}C_5N$ systems investigated in this work exhibit significantly more promising $H_2$ storage capacities, as summarized in Table 3. Although Li and K are associated with environmental and safety concerns, they remain attractive dopants for hydrogen storage because of their low atomic weights. Their light masses contribute minimally to the overall weight of the host material, resulting in a higher gravimetric hydrogen storage capacity than heavier metals. Furthermore, only six dopant atoms are introduced into the host material, representing a very small fraction of the total system. Consequently, the amount of dopant required is minimal, which reduces potential environmental and safety concerns while retaining the advantages of enhanced hydrogen adsorption.

Table 3. Gravimetric density of $H_2$ compared with different storage systems based on DFT calculations. $N_{TOTAL}$, $E_{ads}$, and $C_T$ represent the number of adsorbed $H_2$, adsorption energy, and gravimetric weight capacity, respectively.

| $H_2$ storage system | $N_{TOTAL}$ | $E_{ads}$ (eV) | $C_T$ (wt%) |
| --- | --- | --- | --- |
| $6Li\text{-}C_5N$ (this work) [a] | 48 | -0.17 | 9.42 |

| 6Na-$C_5N$ (this work) [a] | 48 | -0.17 | 8.61 |
|---|---|---|---|
| 6K-$C_5N$ (this work) [a] | 48 | -0.17 | 7.93 |
| 6Sc-$C_5N$ [41][a] | 36 | -0.22 | 5.81 |
| 6Ti-$C_5N$ [41][a] | 36 | -0.22 | 5.73 |
| 6V-$C_5N$ [41][a] | 36 | -0.17 | 5.65 |
| 2Ti-AzaCOF [68][b] | 42 | -0.43 | 9.34 |
| Li-BeS [69][b] | 54 | -0.24 | 6.51 |
| 4Ca-$C_3N_2$ [37] [a] | 22 | -0.24 | 7.53 |
| 4Mg-$C_3N_2$[37] [a] | 25 | -0.24 | 9.47 |
| 4K-$C_3N_2$ [37] [a] | 24 | -0.19 | 8.21 |
| 4Li-BO [65][a] | 16 | -0.21 | 11.75 |
| 4Na-BO [65][a] | 16 | -0.18 | 9.52 |
| 4K-BO [65][a] | 20 | 0.19 | 9.80 |
| 4Ca-BO [65][a] | 24 | -0.20 | 11.43 |
| Mg-g-$C_6N_7$ [51] [a] | 10 | -0.18 | 10 |
| Mg-g-$C_2N$ [63][a] | 30 | -0.12 | 6.79 |
| 2Li-BCN [52] [a] | 16 | -0.12 | 10.10 |
| 2Na-BCN [52] [a] | 16 | -0.19 | 9.18 |
| 2Mg-BCN [52] [a] | 16 | -0.13 | 9.11 |
| 2K-BCN [52] [a] | 16 | -0.24 | 8.41 |
| 2Ca-BCN [52] [a] | 16 | -0.28 | 8.36 |
| Ti-$C_2N$ [70] [a] | 10 | -0.28 | 6.81 |
| Y-g-$C_3N_2$ [71][b] | 9 | -0.33 | 8.55 |
| $C_3N_3$ [72][c] | 15 | -0.25 | 16.13 |
| Li-3D-$B_2P_2$ [59][a] | 60 | -0.15 | 7.07 |
| Na-3D-$B_2P_2$ [59][a] | 60 | -0.11 | 6.36 |
| Li-MOF-5 [73] [b] | 18 | - | 4.30 |
| Sc-g-$C_3N_4$ [74][b] | 7 | -0.39 | 8.55 |
| Li-$B_4N$ [64][b] | 16 | -0.16 | 6.23 |
| Li-$BeN_4$ [75][a] | 7 | -0.22 | 8.32 |
| Ti-graphene [76][b] | 8 | -0.21 | 6.30 |
| Sc-SnC, Ti-SnC, Pd-SnC, Cu-SnC, Ag-SnC [77][c] | - | -0.24 to -0.53 | 5.5 |

[a] GGA-PBE/DFT-D3

[b] GGA-PBE/DFT-D2

[c] VDW-DF/DZP (double zeta polarised)

It is equally important to consider the $C_5N$ and its volumetric capacity for $H_2$ storage. The total volume of $C_5N$ considered in this study is given by Eq.(10)

$$V_{unit\ cell} = a * b * c * sin 60^0 \quad (10)$$

Where a=b=20.93 Å, 'c' is given by $d+d_{up}+d_{down}$. The d refers to the thickness of atomic layers as shown in Figure S6, $d_{up}$ and $d_{down}$ are van der Waals radii of the atoms on its sides.

Table 4. Volumetric $H_2$ storage capacities calculation as per the simulated results.

| System | Planar area =(a*b) (Å$^2$) | $c=d+d_{up}+d_{down}$ | | | $V_{unit\ cell}$ | |
|---|---|---|---|---|---|---|
| | | d (Å)* | $d_{up}$ (Å) | $d_{down}$ (Å) | (Å$^3$) | in (L) |
| 6Li-48$H_2$-$C_5N$ | 438.06 | 7.76 | 1.2 | 1.2 | 3854.41 | 3.85E-24 |
| 6Na-48$H_2$-$C_5N$ | 438.06 | 8.90 | 1.2 | 1.2 | 4286.90 | 4.29E-24 |
| 6K-48$H_2$-$C_5N$ | 438.06 | 9.27 | 1.2 | 1.2 | 4427.17 | 4.43E-24 |

Since 48$H_2$ molecules were adsorbed on 6M-$C_5N$ monolayers, a mass of 1.61E-22 g was calculated using molar mass and Avogadro's number of 6.022E23 molecules/mol. The density of $H_2$ stored in 6Li-, 6Na-, and 6K-$C_5N$ was obtained as 41.74, 37.53, and 36.34 g/L, respectively. The calculated volumetric $H_2$ storage capacities indicate the competitive performance where the Li system surpasses the DoE benchmark of 40 g/L, while the Na- and K-systems approach near the benchmark standards, suggesting promising material for $H_2$ storage.

The $E_b$ and $E_{ads}$ plots, desorption temperatures, extraction of DOS/PDOS plot data, recovery time calculations, and thermodynamic analyses were performed using the data provided in Tables S1 and Table S2 of the Shomate Equation to get the chemical gas potential of $H_2$ molecule, together with the Python code included in the Supporting Information section.

## 4. Conclusion

In this study, an organic $C_5N$ monolayer was investigated as a potential $H_2$ storage material via decoration with alkali metals (Li, Na, and K). The $C_5N$ can stably accommodate up to six dopants (6M-$C_5N$; M= Li, Na, K), with average binding energies of -2.40, -1.79, and -1.90 eV per dopant for Li, Na, and K, respectively. These values are significantly stronger than those $E_c$ of the corresponding bulk metals, effectively suppressing dopant aggregation and ensuring uniform dispersion. Furthermore, AIMD simulations confirm the thermal and structural stability of the 6M-$C_5N$. $H_2$ adsorption analysis reveals that 6M-$C_5N$ can accommodate up to 48$H_2$ molecules, yielding significantly high gravimetric storage capacities of 9.42, 8.61, and 7.93 wt% for 6Li-, 6Na-, and 6K-$C_5N$, respectively. These values exceed the targets set by the U.S. DoE. In addition, the calculated volumetric capacities are 41.74, 37.53, and 36.34 g/L. The reversibility analysis indicates that $H_2$ adsorption is favoured at pressures above 10 bar and temperatures between 280 and 320 K, while desorption occurs at pressures below 10 bar and at 320 K, suggesting practical tunability under operating conditions.

## Acknowledgments

The authors sincerely acknowledge the University of New England for providing the PhD scholarship that supported this research. This research was supported by computational resources provided by the Australian Government through the National Computational Infrastructure (NCI Australia) and the Pawsey Supercomputing Research Centre under NCMAS Merit Allocation. The authors also acknowledge funding from the ARC Training Centre for the Global Hydrogen Economy (GlobH2E, IC200100023).